\documentstyle[aps,eqsecnum,preprint]{revtex}
\begin{document}
\newcommand{\1}{$\clubsuit$}
\newcommand{\2}{$\spadesuit$}
\tighten
\draft
\title{Best Unbiased Estimates for the Microwave Background Anisotropies}

\author{L. P. Grishchuk\thanks{e-mail: grishchuk@astro.cf.ac.uk}}
\address{Department of Physics and Astronomy, University of Wales Cardiff, \\
Cardiff CF2 3YB, United Kingdom \\ and \\ Sternberg Astronomical 
Institute, Moscow University, \\ 119899 Moscow V-234, Russia.}

\author{J\'er\^ome Martin\thanks{e-mail: jmartin@ccr.jussieu.fr}}
\address{Laboratoire de Gravitation et Cosmologie Relativistes, \\ 
Universit\'e Pierre et Marie Curie, CNRS/URA 769, \\
Tour 22/12, Boite courrier 142, 4, place Jussieu, \\
75252 Paris, France.}

\date{\today}
\maketitle

\begin{abstract}
It is likely that the observed distribution of the microwave background
temperature over the sky is only one realization of the underlying random
process associated with cosmological perturbations of quantum-mechanical
origin. If so, one needs to derive the parameters of the random process,
as accurately as possible, from the data of a single map. These parameters 
are of the utmost importance, since our knowledge of them would help us to 
reconstruct the dynamical evolution of the very early Universe.
It appears that the lack of ergodicity of a random process on a 2-sphere   
does not allow us to do this with arbitrarily high accuracy. We are left 
with the problem of finding the best unbiased estimators of the participating 
parameters. A detailed solution to this problem is presented in this article. 
The theoretical error bars for the best unbiased estimates are derived 
and discussed. 
\end{abstract}



\section{Introduction}

The existing and planned measurements of the cosmic 
microwave background (CMB) anisotropies belong to the category of 
astronomical observations which promise a direct link
to fundamental physics, and therefore they attract additional attention.   
\par
The mere detection of the quadrupole anisotropy at
the level $\delta T/T = 5\cdot 10^{-6}$ \cite{Smooth} allows us to conclude that the
Universe remains to be homogeneous and isotropic (all dimensionless deviations 
are smaller than 1) at scales much larger than the present-day Hubble 
radius $l_H$ and up to distances about 500 times 
longer than $l_H$ \cite{GZ}. The significance of this 
result lies in the fact that such large scales are not directly 
observable now and will be accessible only to astronomers of a very remote 
future. At still longer scales, the homogeneity and isotropy of the 
Universe cannot be guaranteed, in the sense that some deviations can be 
larger than 1 without conflicting the CMB observations \cite{GZ}. The 
transition 
from spatially flat cosmological models to open models does not
affect these conclusions considerably \cite{GB}.
\par
The mere existence of the long-wavelength cosmological perturbations, 
responsible for the observed large-angular-scale anisotropy, 
requires them to have special phases and to exist at the previous 
radiation-dominated stage in the form of standing, rather than 
travelling, waves \cite{G1}. This conclusion follows from the Einstein 
equations, 
if we trust them to propagate the observed perturbations back in
time up to, at least, the era of primordial nucleosynthesis without destroying
the homogeneity and isotropy of that era. The distribution of phases can 
be only very narrow (highly squeezed) with two peaks separated by $\pi$
(see \cite{G1} and references there) - a sort of the "phase 
bifurcation" \cite{Schleich}.  
\par
The already identified properties of the presently existing long-wavelength
cosmological perturbations raise sharply the issue of their origin. Although
several schemes are logically possible, it was argued \cite{G1} that the 
quantum-mechanical generation of cosmological perturbations was likely to be 
the one
to do the job. If so, not only the existing requirements are satisfied but 
also some new specific consequences follow.
\par
In a broad sense, the quantum-mechanical generation of cosmological 
perturbations means the Schr\"odinger
evolution of the intial vacuum state (no "particles"-perturbations) into the 
present-day multi-particle state (many "particles"-perturbations). Even the
simplest, linear and quadratic, interaction Hamiltonians are capable of
producing a variety of multi-particle states: coherent states (result of 
the action of a force, linear Hamiltonians), squeezed vacuum states (result
of the parametric influence, quadratic Hamiltonians), squeezed coherent 
states (combination of the two above).    
All these states have Gaussian wave-functions and are in this sense 
Gaussian. A single word "Gaussian" is too general to distinguish 
between these states. The difference between them lies 
in the statements regarding the mean values (zero - for squeezed vacuum 
states, nonzero - for the rest of the states) and the variances (equal - for
coherent states, nonequal - for other states) of the conjugate variables
characterizing the state, such as generalized coordinates and momenta, or 
quadrature components of the field, or, loosely, 
amplitude and phase, etc. In case of cosmological perturbations, 
there is no natural and unavoidable mechanism for 
generation of coherent states, but there is such one for generation of
squeezed vacuum states: parametric (superadiabatic) interaction of the
quantized perturbations with strong variable gravitational field of the
very early Universe (see \cite{G1} and references therein). The fact that 
the cosmological perturbations are being generated 
specifically in the squeezed vacuum quantum states (and not, say, in the
"most classical" coherent states) dictates a number of properties of 
the perturbations themselves and the CMB anisotropies caused by
them \cite{G2}. Squeezing is a physical phenomenon, not a formalism or a
language. 
\par      
Let us imagine that the accurately measured distribution of the CMB 
temperature over the sky is decomposed over spherical harmonics (for
precise definitions see Section 2):  
\begin{eqnarray}
\label{Rcfond}
\frac{{\rm \delta }T}{T}(\theta ,\varphi ) 
&=& \sum _{l=0}^{\infty }\sum _{m=-l}^{+l}c_{lm}Y_{lm}(\theta ,\varphi ).
\nonumber
\end{eqnarray}
A priori, there may be nothing inherently random or quantum-mechanical behind
this observed distribution of the cosmic temperature, in the same sense in 
which there is nothing inherently random or  
quantum-mechanical behind, say, the observed distribution of the 
stars of our Galaxy over the sky. If the measured CMB temperature map 
is a reflection of a 
given classical distribution of matter and gravitational fields, 
the derived fixed numbers $c_{lm}$ is about all what we can 
extract from the map. The notion of the random temperature
arises if the cosmological perturbations responsible for the anisotropies
were randomly generated. We should then interpret a particular perturbation
field and the observed numbers $c_{lm}$ caused by this field as one 
specific realization of the random process. The objective then is
to find out and characterize the underlying random process, as accurately
as possible, using the data of a single observed map.   
\par
We maintain the view that the cosmological perturbations responsible for the
observed anisotropies were generated quantum-mechanically. If they could be
generated in coherent states, the mean quantum-mechanical
values of the perturbation field would be nonzero, and they would be 
surrounded by small (at the level of the zero-point quantum oscillations) 
Gaussian fluctuations. Correspondingly, the
theoretical distributions for the $c_{lm}$ coefficients would have had nonzero
means with small Gaussian fluctuations around them. Roughly speaking,
in this case, the mean numerical values of $\delta T/ T$ would be 
determined 
by the mean values of the perturbation field. In contrast, in case of the 
squeezed vacuum states, the mean values of the perturbation field and 
the mean values of the $c_{lm}$ coefficients are zero, but the Gaussian 
fluctuations around them are large. Roughly speaking, in this case, the
numerical values of $\delta T/ T$ are determined by the dispersion
(square root of variance) of the perturbation field. In both cases, the
words about Gaussian distributions should be taken with a great care.  
Whatever we are able to calculate presently, relies on the assumption that 
cosmological perturbations are weak and that the absolute value of 
$\delta T/ T$ is a small, less than 1, number. On the other hand,
if the variable $\delta T/ T$ obeys a Gaussian (normal) distribution
law, this quantity may take, even though with a small probability 
density, arbitrarily large values, which is in conflict with our 
initial assumption. In addition, the short wavelength perturbations
became nonlinear in the course of evolution, and many extra physical 
processes were involved in the producing of 
the small-angular-scale anisotropies. However, in this paper, we will ignore 
these difficulties and will work with exact normal zero-mean  
distributions (Sections 2, 3).     
\par
In case of squeezed vacuum quantum states, and the corresponding normal 
zero-mean distributions for the $c_{lm}$ coefficients, the underlying random 
process is completely characterized by
the set of variances $\sigma_l^2$  $(l = 0, 1, 2, 3,...)$ of the $c_{lm}$
distributions. These quantities are calculable if the cosmological model
is postulated and, vice versa, the cosmological model can be determined
if these quantities are known from observations. Specifically, 
the quantities $\sigma_l^2$ are calculable if the time dependence 
of the cosmological scale factor describing the very early Universe is
chosen, and assuming that the rest of the evolution is known. 
Moreover, for simple cosmological models, the 
quantities $\sigma_l^2$ are related to each other and are all 
expressible through a small number of parameters. 
This happens, for example,
if one assumes that the scale factor of the very early Universe 
had obeyed one of the power-law time dependences.  
In these cases, the problem of extracting cosmological
information from a given map simplifies and reduces to the problem of
determining those few parameters and testing those models. However, 
the real situation with the data can be more complicated \cite{Gorski}. The 
observer may not be willing to believe any of theoretical cosmological
models, even if he/she accepts the view that the underlying distributions 
should be normal zero-mean distributions. A more ambitious task than
testing each of many possible models is 
the derivation of the true behaviour of the very early Universe from 
the observations, that is, from the observed set of numbers $\sigma_l^2$.   
This is the position that we adopt in this paper. We begin from
observations rather than from theory. Concretely, 
we want to answer the following question: What can be stated about the set 
of independent $\sigma_l^2$ on the grounds of a single observed map ?    
\par
We assume that the sky coverage is complete, the foreground sources and 
contaminating signals are under full control, the instrumental errors 
are negligibly small, the angular resolution is arbitrarily high, and the map
is constructed from the raw data in the most intelligent and effective
way \cite{Wright,Teg}. At the first sight, under these conditions, an 
arbitrarily 
accurate determination of all the $\sigma_l^2$ should not 
present a big problem. Indeed, the angular correlation function 
$K(\delta)$, constructed as a result of averaging over many maps 
("many universes"), has the form    

\begin{eqnarray} 
\label{varz}
\langle\frac{{\rm \delta }T}{T}(\vec{e}_1)
\frac{{\rm \delta }T}{T}(\vec{e}_2)\rangle  = K(\delta) =
\frac{1}{4\pi }\sum _{l=0}^{\infty }\sigma _l^2(2l+1)P_l(\cos \delta ).
\nonumber
\end{eqnarray}

If the function $K(\delta)$ is known, each of the $\sigma_l^2$ can be 
easily obtained from it. One simply needs to integrate over $\delta$ 
the product of $K(\delta)$ with the respective Legendre 
polynomial $P_l(\cos \delta)$. It is true that we have access to only one 
realization of the random process, not to infinitely many implied in 
the construction of the $K(\delta)$, but this is not an 
obstacle by itself. If the process is ergodic \cite{Ya,Pap} (see Section 4), 
the correlation functions can be built from a single realization, and the 
parameters of the random process, such as $\sigma_l^2$, can be 
determined with arbitrarily high accuracy. What is required for the 
ergodicity of a random process in time or in 3-space is the 
decay of correlations in the limit of very large temporal
or spatial separations. Then, the ensemble averages can be replaced by
integrals over time or 3-space, and the parameters derived from one 
realization are true parameters 
with probabilty 1. In theoretical cosmology, the long-distance behaviour 
of the perturbation field is partially in our hands. It can be assumed to be 
appropriate, so that the field in 3-space can be made ergodic: the access to
one realization would be sufficient for the determination of all parameters
of the perturbation field. The difficulty apparently comes about when the 
3-dimensional ergodic process is being reduced to a 2-dimensional random 
process on a sphere, what effectively takes
place when cosmological perturbations produce anisotropies in the temperature
distribution over the sky. It appears that a random process on a 2-sphere
can never be ergodic (in the sense of replacement of the ensemble averages
by the integrations over the sphere) and the $\sigma_l^2$ can never be 
determined with arbitrarily high accuracy from a single map (Section 4). 
Possibly, this is a statement well known to mathematicians but we could 
not find an adequate reference.        
\par
Without being able to find the true values of $\sigma_l^2$ 
we are left with the problem of estimating these parameters as accurately 
as possible. One can imagine that one is given a very precise map of the
CMB sky, but the question is what to do with this map. This is a problem 
from the quite well developed theory of estimation and statistical inference 
(see, for example \cite{Meyer,Sage}). To evaluate a parameter we need 
to build an estimator - a random variable constructed from the 
original random process. The estimator is "unbiased" if its expectation
value is equal to the true value of the parameter. And
the estimator is the "best" if its variance is minimal among all possible
estimators. In Section 5, we find the best unbiased estimator 
for $\sigma_l^2$, and in Section 6 we find the best unbiased
estimator for $K(\delta)$. 
\par
A concrete numerical value of the best 
unbiased estimator acquired at the observed map is the best unbiased 
estimate of the corresponding parameter. Usually, one is not satisfied
with the best estimate alone, but wants to surround the estimate by 
appropriate error bars. This requires new definitions
and criteria (see Section 7). Approximately, but not exactly, the size 
of the error box is characterized by the variance of the
best unbiased estimator. Not surprisingly, since we have limited our
discussion to normal zero-mean distributions, the most "natural" 
estimators turn out to be also the best unbiased estimators, and 
the maximum likelihood estimators, etc. In this way, this paper makes 
contact with previous work on the subject \cite{early}.        
\par
Before concluding the Introduction we need to make two comments.
\par
First, our analysis is built on the assumptions that the observed map is 
only one realization of a random process, and that the underlying 
distributions are normal zero-mean distributions. These assumptions 
are in fact consequences of the parametric quantum-mechanical generating 
mechanism of the perturbations, but they are also testable hypotheses on 
their own. Logically, one needs first to show that one is 
dealing with a random process before trying to find out its 
characteristics. If we were experimenting with a noisy voltage generator, 
we could compare several sufficiently long records in order to argue 
that we were dealing with the different realizations of one and
the same random process. In cosmology, we have control over 
only one record, only one CMB map. The assumptions made above can 
hardly be proven rigorously, but one can possibly find evidence in
their support. Alternatively, they can be disproved at some 
level of confidence. One possibility to test these assumptions 
was indicated in Ref. \cite{G2}. The product of two random variables
${\rm \delta }T/T(\vec{e}_1)$ and ${\rm \delta }T/T(\vec{e}_2)$ is
a new random variable $v$. The probability density function (p.d.f.) for 
$v$ was derived, and its functional form was shown to be quite special [6].   
The mean value of $v$ is $K(\delta)$ but the variable $v$ 
is not supposed to be a good estimator of $K(\delta)$ since the variance of
$v$ is very big. However, this big variance should be present in the 
original observational data (before any angular integrations over the map
are performed) 
if our statistical assumptions are correct. It looks unlikely that the
quite special functional form of the p.d.f. for $v$ can be mimicked by the 
anisotropies caused by perturbations of any other origin. In view of the 
forthcoming massive "pixelization" of the CMB sky, it will probably 
be possible to test directly whether the map values of the $v$ satisfy, 
at least approximately, the theoretically derived p.d.f. for $v$. 
In this paper, we effectively assume that this is the case.       
\par
Second, in our analysis we do not need to specify the nature of the 
cosmological perturbations responsible for the observed large-angular-scale
anisotropy. They can be density perturbations, or rotational perturbations,
or gravitational waves. However, in order to build a correct general 
picture, it is very important to know what kind of perturbations we 
are actually dealing with. Hopefully, this question can be answered in
future observations, with the help, for example, of polarization
measurements (for a recent paper on the subject, see \cite{KKS}). 
So far, one can only rely on the theory. The quantum-mechanical
generating mechanism, originally developed for gravitational waves, can also
be applied, under certain conditions, to density perturbations and 
rotational perturbations. The contribution of each type of perturbations
to the $\delta T /T$ can be calculated. According to the calculations of
Ref.\cite{G3}, if the observed large-angular-scale anisotropies are 
indeed caused by cosmological perturbations of quantum-mechanical origin, 
gravitational waves are at least as important as density perturbations
and provide a somewhat larger contribution than density perturbations.

\section{Various representations for ${\rm \delta }T/T$}

The quantity which appears naturally in the theory of the CMB anisotropies
is a relative variation of the temperature seen in a given direction
$\vec{e}$ on the sky: ${\rm \delta }T/T(\vec{e})$. This quantity is a 
function of the angular coordinates on the celestial sphere
\begin{equation}
\label{defz}
\frac{{\rm \delta }T}{T}(\vec{e}) \equiv \frac{{\rm \delta }T}{T}
(\theta ,\varphi ). \nonumber
\end{equation}
\par
It is convenient to define three different representations for this function:
\begin{eqnarray}
\label{Ra}
\frac{{\rm \delta }T}{T}(\theta ,\varphi ) &=
& \sum _{l=0}^{\infty }\sum _{m=-l}^{+l} 
[a_{lm}Y_{lm}(\theta ,\varphi )+a_{lm}^*
Y_{lm}^*(\theta ,\varphi )], \\
\label{Rb}
&=& \sum _{l=0}^{\infty }\biggl(b_{l0}^c\frac{Y_{l0}^c(\theta ,\varphi )}
{\sqrt{2}}+\sum _{m=1}^l[b_{lm}^cY_{lm}^c(\theta ,\varphi )+
b_{lm}^sY_{lm}^s(\theta ,\varphi)]\biggr), \\
\label{Rc}
&=& \sum _{l=0}^{\infty }\sum _{m=-l}^{+l}c_{lm}Y_{lm}(\theta ,\varphi ),
\end{eqnarray}
where the orthonormal spherical harmonics, either complex $Y_{lm}$,
or real $Y_{l0}^c/\sqrt{2}$, $Y_{lm}^c$ and $Y_{lm}^s$, are described in 
the Appendix. In what follows, the index $l$ runs from $0$ to 
$\infty $, and the index $m$ runs either between $-l$ and $+l$ 
or between $1$ and $l$, as will be explicitely specified. 
\par
Although the transitions between Eqs. (\ref{Ra}), (\ref{Rb}), and (\ref{Rc}) 
are quite
straightforward, each of the representations has its own advantages 
and we will use them below. Since every factor $\sqrt 2$ can eventually 
prove to be very important,
we need a rigorous, even if somewhat pedantic, description of these 
representations. The coefficients $a_{lm}$ are complex and can be 
expressed in terms of their real and imaginary parts: 
$a_{lm}=a_{lm}^r+ia_{lm}^i$. In this a-representation, the function
${\rm \delta }T/T(\vec{e})$ is manifestely real, but the number of the
a-coefficients is larger than necessary. The b-representation is a little
cumbersome, but it has to be regarded as canonical in the sense that it
contains only real and independent coefficients $b_{lm}^A$ ($A=c,s$) 
and does not require any extra constraints. The properties of other 
representations will be derived from the properties of this one. 
Finally, the coefficients $c_{lm}$ of the c-representation are complex, 
and in order to have real ${\rm \delta }T/T(\vec{e})$, they must 
satisfy the relationship $c_{lm}^*=c_{l,-m}$. They can be also written 
as $c_{lm}=c_{lm}^r+ic_{lm}^i$. Then, the last relationship implies 
$c_{lm}^r=c_{l,-m}^r$ and $c_{lm}^i=-c_{l,-m}^i$. 
\par
Let us now describe how these representations are related to each other.
\par
The relationship between the b-representation and the a-representation is 
expressed through the following equations
\begin{eqnarray}
\label{linkba1}
b_{l0}^c &=& 2a_{l0}^r, \\
\label{linkab2}
b_{lm}^c &=& \sqrt{2}(a_{lm}^r+a_{l,-m}^r) \quad m\ge 1, \\
\label{linkba3}
b_{lm}^s &=& -\sqrt{2}(a_{lm}^i-a_{l,-m}^i) \quad m\ge 1.
\end{eqnarray}
The link between the c-representation and the a-representation is given by
\begin{eqnarray}
\label{linkca}
c_{lm}=a_{lm}+a_{l,-m}^*.
\nonumber
\end{eqnarray}
The link between the c-representation and the b-representation can be 
written as
\begin{eqnarray}
\label{linkcb1}
c_{l0} &=& b_{l0}^c, \\
\label{linkcb2}
c_{lm} &=& \frac{1}{\sqrt{2}}(b_{lm}^c-ib_{lm}^s) \quad m\ge 1, \\
\label{linkcb3}
c_{lm} &=& \frac{1}{\sqrt{2}}(b_{l,-m}^c+ib_{l,-m}^s) \quad m\le -1.
\end{eqnarray} 
Using the previous equations it is easy to check that 
$c_{l,-m}=1/\sqrt{2}(b_{lm}^c+ib_{lm}^s)=c_{lm}^*$. 
We can also express the real and imaginary parts of the 
coefficients $c_{lm}$ in terms of the coefficients $b_{lm}^A$. 
Explicitly we have 
\begin{eqnarray} 
\label{linkcb4} 
c_{l0}^r &=& b_{l0}^c, \quad c_{l0}^i=0, \nonumber \\ 
\label{linkcb5}
c_{lm}^r &=& \frac{b_{lm}^c}{\sqrt{2}} \quad m\ge 1, \nonumber \\
\label{linkcb6}
c_{lm}^i &=& -\frac{b_{lm}^s}{\sqrt{2}} \quad m\ge 1. \nonumber
\end{eqnarray}
If $m\le -1$, then $b_{lm}^c$ and $b_{lm}^s$ have to be replaced 
by $b_{l,-m}^c$ and $b_{l,-m}^s$, respectively. Finally, 
Eqs. (\ref{linkcb1})-(\ref{linkcb3}) 
can be inverted, and one can express the coefficients $b_{lm}^A$ in terms 
of the coefficients $c_{lm}$. One obtains  
\begin{eqnarray}
\label{linkbc1}
b_{l0}^c &=& c_{l0}, \\
\label{linkbc2}
b_{lm}^c &=& \frac{1}{\sqrt{2}}(c_{lm}+c_{lm}^*) \quad m\ge 1, \\
\label{linkbc3}
b_{lm}^s &=& \frac{i}{\sqrt{2}}(c_{lm}-c_{lm}^*) \quad m\ge 1.
\end{eqnarray}

\section{The probability density functions, mean values, and variances}
 
We need to formulate our statistical assumptions about the CMB temperature.
We will start from the b-representation. Our assumptions are
as follows: 1) all the coefficients $b_{l0}^c$, $b_{lm}^c$ and $b_{lm}^s$ 
are statistically independent random variables, 2) each individual
variable is normally distributed with a zero mean, 3) all variables
with the same index $l$ have the same standard deviation $\sigma_l$. 
In other words, the probability density functions for the b-coefficients 
are given by the expressions
\begin{eqnarray}
\label{pdfb1}
f(b_{l0}^c) &=
& \frac{1}{\sqrt{2\pi }\sigma _l}e^{-\frac{(b_{l0}^c)^2}{2\sigma _l^2}}, \\
\label{pdfb2}
f(b_{lm}^c) &=
& \frac{1}{\sqrt{2\pi }\sigma _l}e^{-\frac{(b_{lm}^c)^2}{2\sigma _l^2}} 
\quad m\ge 1, \\
\label{pdfb3}
f(b_{lm}^s) &=
& \frac{1}{\sqrt{2\pi }\sigma _l}e^{-\frac{(b_{lm}^s)^2}{2\sigma _l^2}} 
\quad m\ge 1.
\end{eqnarray}
The $\delta T/ T$ taken in a given direction is a random variable, while
the $\delta T/ T$ treated as a function of $\theta, \varphi$ is a 
random (stochastic) process. 
\par
Having the p.d.f.'s one can calculate various useful expectation values.
For the mean values of the $b$-coefficients, one obtains  
\begin{eqnarray}
\label{meanb}
\langle b_{l0}^c\rangle=\langle b_{lm}^c\rangle=\langle b_{lm}^s\rangle=0 
\quad m\ge 1. \nonumber 
\end{eqnarray}
For the quadratic combinations, the result is
\begin{eqnarray}
\label{quadrab1}
\langle b_{l_10}^cb_{l_20}^c\rangle &=& \sigma _{l_1}^2\delta _{l_1l_2}, \\
\label{quadrab2}
\langle b_{l_1m_1}^cb_{l_2m_2}^c\rangle &=& \langle b_{l_1m_1}^sb_{l_2m_2}^s
\rangle=
\sigma _{l_1}^2\delta _{l_1l_2}\delta _{m_1m_2}  \quad m_1,m_2 \ge 1, 
\nonumber \\
\label{quadrab3}
\langle b_{l_1m_1}^cb_{l_2m_2}^s\rangle &=& \langle b_{l_1m_1}^sb_{l_2m_2}^c
\rangle=0 
\quad m_1,m_2 \ge 1. \nonumber
\end{eqnarray}
In a similar manner, one can also determine the quartic combinations. 
\par
We can now deduce the statistical properties of the two other 
representations. For the a-representation, all the coefficients 
$a_{lm}^r$ and $a_{lm}^i$ can be taken as statistically independent, and 
their p.d.f.'s can be written as
\begin{eqnarray}
\label{pdfa}
f(a_{lm}^r) = \frac{2}{\sqrt{2\pi }\sigma _l}e^{-\frac{2(a_{lm}^r)^2}
{\sigma _l^2}}, \quad
f(a_{lm}^i) = \frac{2}{\sqrt{2\pi }\sigma _l}e^{-\frac{2(a_{lm}^i)^2}
{\sigma _l^2}}. \nonumber
\end{eqnarray}
To find the expectation values one can use these p.d.f.'s, and can also
make a consistency check with the help of equations 
(\ref{linkba1})-(\ref{linkba3}) and (\ref{pdfb1})-(\ref{pdfb3}).   
The mean values of $a_{lm}^r$ and $a_{lm}^i$ are obviously zero (and 
therefore this is also the case for $a_{lm}$ and $a_{lm}^*$):
\begin{eqnarray}
\label{meana}
\langle a_{lm}^r\rangle=\langle a_{lm}^i\rangle=0. \nonumber
\end{eqnarray}
The quadratic combinations of $a_{lm}^r$ and $a_{lm}^i$ have the values 
\begin{eqnarray}
\label{quadraa1}
\langle a_{l_1m_1}^ra_{l_2m_2}^r\rangle &=& \langle a_{l_1m_1}^ia_{l_2m_2}^i
\rangle=
\frac{\sigma _{l_1}^2}{4}\delta _{l_1l_2}\delta _{m_1m_2}, \nonumber \\
\label{quadraa2}
\langle a_{l_1m_1}^ra_{l_2m_2}^i\rangle &=& \langle a_{l_1m_1}^ia_{l_2m_2}^r
\rangle=0, \nonumber
\end{eqnarray}
from which one derives 
\begin{eqnarray}
\label{quadraa3}
\langle a_{l_1m_1}a_{l_2m_2}\rangle=0, 
\qquad \langle a_{l_1m_1}a_{l_2m_2}^*\rangle=
\frac{\sigma _{l_1}^2}{2}\delta _{l_1l_2}\delta _{m_1m_2}. \nonumber
\end{eqnarray}
Some of nonvanishing quartic combinations are given by the expressions
\begin{eqnarray}
\label{quartic1}
\langle a_{l_1m_1}^ra_{l_2m_2}^ra_{l_3m_3}^ra_{l_4m_4}^r\rangle &=
& \frac{\sigma _{l_1}^2\sigma _{l_3}^2}{16}\delta _{l_1l_2}
\delta _{m_1m_2}\delta _{l_3l_4}\delta _{m_3m_4}
+\frac{\sigma _{l_1}^2\sigma _{l_2}^2}{16}\delta _{l_1l_3}
\delta _{m_1m_3}\delta _{l_2l_4}\delta _{m_2m_4} \nonumber \\
& & +\frac{\sigma _{l_1}^2\sigma _{l_2}^2}{16}\delta _{l_1l_4}
\delta _{m_1m_4}\delta _{l_2l_3}\delta _{m_2m_3}, \nonumber \\
\label{quartic2}
\langle a_{l_1m_1}^ra_{l_2m_2}^ra_{l_3m_3}^ia_{l_4m_4}^i\rangle &=&
\frac{\sigma _{l_1}^2\sigma _{l_3}^2}{16}\delta _{l_1l_2}
\delta _{m_1m_2}\delta _{l_3l_4}\delta _{m_3m_4}, \nonumber \\
\label{quartic3}
\langle a_{l_1m_1}a_{l_2m_2}a_{l_3m_3}^*a_{l_4m_4}^*\rangle &=&
\frac{\sigma _{l_1}^2\sigma _{l_2}^2}{4}(
\delta _{l_1l_3}\delta _{m_1m_3}\delta _{l_2l_4}\delta _{m_2m_4}+
\delta _{l_1l_4}\delta _{m_1m_4}\delta _{l_2l_3}\delta _{m_2m_3}),  
\nonumber
\end{eqnarray}
and other nonvanishing combinations can be obtained by permutting the 
indices $r$, $i$. 
\par
Finally, starting from the postulated distributions in the b-representation,
one can also establish the corresponding equations for the c-representation. 
All the coefficients $c_{l0}^r$, $c_{lm}^r$ and $c_{lm}^i$ ($m>0$) should be 
statistically independent, and their p.d.f.'s  should be written as
\begin{eqnarray}
\label{pdfc1}
f(c_{l0}^r) = \frac{1}{\sqrt{2\pi }\sigma _l}e^{-\frac{(c_{l0}^r)^2}
{2\sigma _l^2}}  \nonumber
\end{eqnarray}
and
\begin{eqnarray}
\label{pdfc2}
f(c_{lm}^r) =\frac{1}{\sqrt{\pi }\sigma _l}e^{-\frac{(c_{lm}^r)^2}
{\sigma _l^2}}, \quad f(c_{lm}^i) = \frac{1}{\sqrt{\pi }\sigma _l}e^
{-\frac{(c_{lm}^i)^2}{\sigma _l^2}}, \quad m \ge 1. \nonumber
\end{eqnarray}
Note that the standard deviation for $c_{l0}^r$ is different from
that for $c_{lm}^r$ and $c_{lm}^i$. Obviously, the mean values of the 
coefficients $c_{l0}^r$, $c_{lm}^r$ and $c_{lm}^i$ vanish
\begin{eqnarray}
\label{meanc}
\langle c_{l0}^r\rangle=\langle c_{lm}^r\rangle=\langle c_{lm}^i\rangle=0. 
\nonumber
\end{eqnarray}
The mean value of the quadratic combination of $c_{l0}^r$ is given by
\begin{eqnarray}
\label{quadrac1}
\langle c_{l_10}^rc_{l_20}^r\rangle = \sigma _{l_1}^2\delta _{l_1l_2}. 
\nonumber
\end{eqnarray}
For other coefficients (where $m_1$ and $m_2$ are not both equal to
zero) one obtains  
\begin{eqnarray}
\label{quadrac2}
\langle c_{l_1m_1}^rc_{l_2m_2}^r\rangle &=
& \frac{\sigma _{l_1}^2}{2}\delta _{l_1l_2}\delta _{m_1m_2}+
\frac{\sigma _{l_1}^2}{2}\delta _{l_1l_2}\delta _{m_1,-m_2}, \nonumber \\
\label{quadrac3}
\langle c_{l_1m_1}^ic_{l_2m_2}^i\rangle &=
& \frac{\sigma _{l_1}^2}{2}\delta _{l_1l_2}\delta _{m_1m_2}-
\frac{\sigma _{l_1}^2}{2}\delta _{l_1l_2}\delta _{m_1,-m_2}, \nonumber \\
\label{quadrac4}
\langle c_{l_1m_1}^rc_{l_2m_2}^i\rangle &=& \langle c_{l_1m_1}^ic_{l_2m_2}^r
\rangle=0. \nonumber
\end{eqnarray}
This leads to  
\begin{equation}
\label{quadrac5}
\langle c_{l_1m_1}c_{l_2m_2}\rangle = 
\sigma _{l_1}^2\delta _{l_1l_2}\delta _{m_1,-m_2}, 
\quad
\langle c_{l_1m_1}c_{l_2m_2}^*\rangle = 
\sigma _{l_1}^2\delta _{l_1l_2}\delta _{m_1m_2}. 
\end{equation}
The last equations are also valid for $m_1=m_2=0$. 
\par
One nonvanishing quartic combination is given by the expression
(others can be obtained by the complex conjugation):
\begin{eqnarray}
\label{quarticc1}
\langle c_{l_1m_1}c_{l_2m_2}c_{l_3m_3}c_{l_4m_4}\rangle &=
& \sigma _{l_1}^2\sigma _{l_2}^2\delta _{l_1l_3}
\delta _{m_1,-m_3}\delta _{l_2l_4}\delta _{m_2,-m_4}+
\sigma _{l_1}^2\sigma _{l_2}^2\delta _{l_1l_4}\delta _{m_1,-m_4}
\delta _{l_2l_3}\delta _{m_2,-m_3} \nonumber \\
&+& \sigma _{l_1}^2\sigma _{l_3}^2\delta _{l_1l_2}\delta _{m_1,-m_2}
\delta _{l_3l_4}\delta _{m_3,-m_4}.
\end{eqnarray}
\par 
Let us now introduce new random variables which will play an important 
role in what follows. Let us define random variables $a_l^2$, 
$b_l^2$, and $c_l^2$ by the equations 
\begin{eqnarray}
\label{defa^2}
a_l^2 &\equiv & \sum _{m=-l}^la_{lm}a_{lm}^*, \nonumber \\
\label{defb^2}
b_l^2 &\equiv & (b_{l0}^c)^2+\sum _{m=1}^l[(b_{lm}^c)^2+(b_{lm}^s)^2], 
\nonumber \\
\label{defc^2}
c_l^2 &\equiv & \sum _{m=-l}^l c_{lm}c_{lm}^*. \nonumber
\end{eqnarray}
Using Eqs. (\ref{linkbc1})-(\ref{linkbc3}) it is easy to 
show that $b_l^2=c_l^2$. 
\par
One can compute the mean values of these new random variables
\begin{eqnarray}
\label{meanabc^2}
2\langle a_l^2\rangle = \langle b_l^2\rangle = \langle c_l^2\rangle = 
(2l+1)\sigma _l^2  \nonumber
\end{eqnarray}
and their variances 
\begin{eqnarray}
\label{vara^2}
\langle a_l^4\rangle-\langle a_l^2\rangle^2 &=& \frac{1}{2l+1}
\langle a_l^2\rangle^2,  \\
\label{varbc^2}
\langle b_l^4\rangle-\langle b_l^2\rangle^2 &=& \langle c_l^4\rangle-
\langle c_l^2\rangle^2 =\frac{2}{2l+1}\langle c_l^2\rangle^2.
\end{eqnarray}
It is important to note that the above relationships are trivial consequences
of the postulated distributions (\ref{pdfb1}) - (\ref{pdfb3}). These 
relationships are always 
true, regardless of what and how is measured, and regardless of whether we have
access to only one realization of the random process (only one sky or portion
of sky) or to 
infinitely many. (But if one wants to use a "cosmic" word, one is free to 
call
the relationships (\ref{vara^2}), (\ref{varbc^2}) the "cosmic variance".)      
\par
The original probability distributions dictate also the p.d.f.'s for 
these quadratic variables. They are the 
so-called $\chi^2$ - distributions. Denoting $\chi^2 = b_l^2/\sigma_l^2$
and $n = 2l+1$ one can write \cite{Meyer}
\begin{eqnarray}
\label{pdfchi}
f(\chi^2, n) = \frac{(\chi^2)^{(n-2)/2} e^{-\chi^2/2}}
{(n/2 - 1)! 2^{n/2}}, \nonumber 
\end{eqnarray}
and for the random variable $b_l^2$ we have 
\begin{equation}
\label{pdfbl2}
f(b_l^2) =  \frac{(b_l^2)^{(n-2)/2} (\sigma_l^2)^{-n/2} e^{-b_l^2/2\sigma_l^2}}
{(n/2 - 1)! 2^{n/2}}
\end{equation} 
\par
So far, we have been concerned with the statistical properties of the
coefficients in the expansion of the random process $\delta T/ T$ 
over the orthonormal spherical harmonics. We can now discuss some properties
of the random process itself. No doubt, these properties follow from the 
properties of the coefficients. Obviously, the process is isotropic in the
sense that the mean value of the $\delta T/ T$ is one and the same number 
(zero) in every direction on the sky 
\begin{equation}
\label{meanz}
\langle\frac{{\rm \delta }T}{T}(\vec{e})\rangle = 0.
\end{equation}
The process is also homogeneous in the sense that 
the angular correlation function depends only on the angle $\delta$
between two directions, but not on directions themselves. For each pair of 
vectors $\vec{e}_1$ and $\vec{e}_2$ separated by the angle $\delta$,
the angular correlation function takes the form 
\begin{eqnarray} 
\label{varz2}
\langle\frac{{\rm \delta }T}{T}(\vec{e}_1)
\frac{{\rm \delta }T}{T}(\vec{e}_2)\rangle = K(\delta) =  
\frac{1}{4\pi }\sum _{l=0}^{\infty }\sigma _l^2(2l+1)P_l(\cos \delta), 
\end{eqnarray}
where $P_l(\cos \delta)$ are the Legendre polynomials.
\par
The three-point correlation function [as well as all correlation functions 
containing an odd number of term ${\rm \delta }T/T(\vec{e}_i)$] vanishes:
\begin{eqnarray}
\label{3z}
\langle\frac{{\rm \delta }T}{T}(\vec{e}_1)\frac{{\rm \delta }T}{T}(\vec{e}_2)
\frac{{\rm \delta }T}{T}(\vec{e}_3)\rangle =0. \nonumber
\end{eqnarray}
The four-point correlation function is given by the expression 
\begin{eqnarray}
\label{4z}
& & \langle\frac{{\rm \delta }T}{T}(\vec{e}_1)\frac{{\rm \delta }T}{T}(\vec{e}_2)
\frac{{\rm \delta }T}{T}(\vec{e}_3)\frac{{\rm \delta }T}{T}(\vec{e}_4)\rangle = 
\frac{1}{(4\pi )^2}\sum _{l=0}^{\infty }\sum _{m=0}^{\infty }
\sigma _l^2\sigma _m^2(2l+1)(2m+1) \nonumber \\
& & \biggl(P_l(\cos \delta _{13})P_m(\cos \delta _{24}) + 
P_l(\cos \delta _{14})P_m(\cos \delta _{23}) + 
P_l(\cos \delta _{12})P_m(\cos \delta _{34})\biggr), \nonumber 
\end{eqnarray}
where the symbol $\delta _{ij}$ denotes the angle between vectors 
$\vec{e}_i$ and $\vec{e}_j$. In a similar manner one can derive the
higher-order correlation functions and express them through the
lower-order ones. 
\par
All the derived expressions are consequences of the postulated distribution
functions (\ref{pdfb1}) - (\ref{pdfb3}). We do not possess a rigorous 
mathematical proof 
of the statement that there exists a one-to-one correspondence between
Eqs. (\ref{pdfb1}) - (\ref{pdfb3}) and the fact that the relevant 
cosmological 
perturbations are placed in the squeezed vacuum quantum states. However, we 
believe this statement is indeed true. At any rate, the quantum-mechanical
expectation values coincide with the corresponding ensemble averages if the
appropriate identifications are made [6]. In particular, the 
quantum-mechanical expression for the angular correlation function 
\begin{eqnarray}
\label{quantumvarz}
\langle 0|\frac{{\rm \delta }T}{T}(\vec{e}_1)\frac{{\rm \delta }T}{T}
(\vec{e}_2)|0\rangle 
=\sum _{l=l_{min}}^{\infty }C_lP_l(\cos \delta) \nonumber 
\end{eqnarray}
coincides with Eq. (\ref{varz2}) if we identify 
\begin{equation}
\label{corresCsig}
\frac{1}{4\pi }(2l+1)\sigma _l^2 = C_l. 
\end{equation}
The quantity $C_l$ explicitely contains the square of the Planck length. This
quantity is calculable when the law of cosmological evolution and the sort of 
cosmological perturbations are specified. The value of the lowest 
multipole $l_{min}$ is also determined by the sort of perturbations. 
In this sense, the abstract
quantities $\sigma_l^2$, which completely characterize the random
process (\ref{defz}), (\ref{pdfb1}) - (\ref{pdfb3}), are also calculable 
and are given by the 
expressions 
\begin{eqnarray}
\label{corresCf}
\sigma _l^2=\frac{4\pi }{2l+1}C_l. \nonumber
\end{eqnarray}

\section{Ergodicity of random processes on a line and on a sphere}

Let us first recall the ergodic theorem \cite{Ya,Pap} for a time-dependent
random process $x(\xi ,t)$ defined on an infinite $t$-line. The symbol
$\xi$ indicates different possible realizations. Let us assume that the
process is stationary, that is, its mean value does not depend on time
\begin{eqnarray}
\label{statio}
\langle x(\xi ,t)\rangle=const=m,  \nonumber
\end{eqnarray}
and its correlation function depends only on the time difference
\begin{eqnarray}
\label{correl}
\langle x(\xi, t + \tau)x(\xi, t)\rangle = B(\tau). \nonumber
\end{eqnarray}
\par
To find the ensemble average of $x(\xi, t)$ at a fixed moment of time, one 
takes a large number $N$ of different realizations and calculates the 
arithmetic mean of the observed values:  
\begin{eqnarray}
\label{deff}
f=\frac{1}{N}\sum _{i=1}^N x(\xi _i,t). \nonumber
\end{eqnarray}
In the limit of $N$ going to infinity, the quantity $f$ tends to the 
theoretical ensemble mean $\langle x(\xi, t)\rangle$ of the random process. 
\par
Let us now consider a situation in which we have access to only one
realization $x(\xi_0, t)$ of the random process. What can we say about
$m$ and $B(\tau)$ on the grounds of this single realization ? The
ergodic theorem defines the conditions under which the ensemble
averages can be replaced by the time averages, so that the $m$ and
$B(\tau)$ can be found from the time integrations of $x(\xi_0, t)$.
\par
Introduce a random variable $x_T(\xi)$ defined by the equation 
\begin{equation}
\label{defxT}
x_T(\xi )=\frac{1}{2T}\int _{-T}^{+T} x(\xi ,t){\rm d}t.
\end{equation}
This variable is an unbiased estimator of $m$ because 
$\langle x_T(\xi )\rangle=m$. However, we can say much more when we take 
the limit
$T\rightarrow \infty$. If the process is such that
\begin{equation}
\label{condB}
\lim _{T\rightarrow \infty }\frac{1}{2T}
\int _{-T}^{+T}B(\tau){\rm d}\tau = 0,  
\end{equation}
then
\begin{eqnarray}
\label{condx}
\lim _{T\rightarrow \infty }\frac{1}{2T}
\int _{-T}^{+T} x(\xi ,t){\rm d}t = m,  \nonumber
\end{eqnarray}
for every realization $\xi$. When condition (\ref{condB}) is satisfied, the 
process 
is called mean-ergodic. The condition (\ref{condB}) can also be expressed as 
the requirement $\sigma _{x_T}^2 \rightarrow 0$ in the limit  
$T\rightarrow \infty$, where 
$\sigma _{x_T}^2$ is the variance of the random variable defined 
by Eq. (\ref{defxT}).
This explains why one is capable of deriving from a single realization 
a true parameter of the ergodic random process (in this case, 
the mean value $m$) with probability 1. Indeed, for every arbitrarily
small $\epsilon$, one has the Tchebysheff inequality 
\begin{eqnarray}
\label{Tcheby}
P(|x_T-m|<\epsilon) \ge 1-\frac{\sigma _{x_T}^2}{\epsilon ^2}, \nonumber
\end{eqnarray}
and the probability goes to 1 when $\sigma _{x_T}^2$ goes to 0.    
A sufficient condition for the validity of Eq. (\ref{condB}) is the vanishing of
the correlations at large temporal separations: 
$\lim _{\tau \rightarrow \infty} B(\tau) = 0$. 
\par
More stringent conditions should be satisfied for the process to be
correlation-ergodic, that is to have 
\begin{equation}
\label{condcorrel}
\lim _{T\rightarrow \infty }\frac{1}{2T}
\int _{-T}^{+T} x(\xi ,t + \tau) x(\xi ,t){\rm d}t = B(\tau),
\end{equation}
for every $\xi$. Here, for simplicity, we will restrict ourselves to 
normal zero-mean ($m=0$) stationary processes. The equality  
(\ref{condcorrel}) allows one 
to find the variance of the process from its single realization. 
This equality takes place if and only if the process 
is such that  
\begin{equation}
\label{condergovar}
\lim _{T\rightarrow \infty }\frac{1}{2T}
\int _{-T}^{+T}|B(\tau)|^2{\rm d}\tau = 0.   
\end{equation}
For normal zero-mean stationary processes, and when the 
condition (\ref{condergovar}) is met, all the higher-order correlation 
functions can 
also be replaced by the time averages. Note that the integrand 
in Eq. (\ref{condergovar}) is a strictly
positive function. The limit of the expression (\ref{condergovar}) is zero 
because the
denominator goes to infinity in the limit $T\rightarrow \infty$. 
\par
We will now try to apply the notions formulated above to a random process on
a sphere. The problem of our interest is the CMB temperature distributed 
over the sky. Strictly speaking, the quantum-mechanically 
generated cosmological 
perturbations form a non-stationary process: the squeezing makes the
temporal correlation function a function of individual moments of time,
and not only of the time difference. This property may turn out to be 
very important for the future observations of short gravitational waves, 
but is irrelevant for our discussion of very long-wavelength 
cosmological perturbations responsible for the microwave background
anisotropies, since the time scale of variations is much much
longer than the interval of time between possible missions for the 
CMB observations. Most importantly, the stationarity - a necessary (but
not sufficient) condition
for a time dependent process to be ergodic - is replaced in our case 
by the analogous properties of isotropy and homogeneity of the process 
on a sphere, see Eqs. (\ref{meanz}), (\ref{varz2}). So, at least the 
necessary conditions 
for our process to be mean-ergodic and correlation-ergodic are satisfied. 
\par
To check the analog of the condition (\ref{condB}), we will use $K(\delta)$ 
instead of $B(\tau)$ and will replace the time integral divided by $T$ 
by the integral over a sphere divided by $4\pi$ - the
surface area of a (unit radius) sphere. The result is
\begin{eqnarray}    
\label{intK}
\frac{1}{4\pi}\int _{S^2}{\rm d}\Omega K(\rm \delta)=
\frac{1}{4\pi}\sigma_0^2. \nonumber
\end{eqnarray}
The right-hand-side of this equation is zero when $\sigma_0^2$ is zero,
that is when the monopole coefficient $b_{00}^c$ in the expansion (\ref{Rb}) is
identically zero, see Eq. (\ref{quadrab1}). In this case, our process 
is indeed mean-ergodic since
the integral over a given map
\begin{eqnarray}   
\label{intz}
\frac {1}{4\pi}\int _{S^2}{\rm d}\Omega\frac{{\rm \delta }T}{T}(\vec{e})= 
\frac{1}{\sqrt{4\pi}}b_{00}^c  \nonumber
\end{eqnarray}
vanishes, and the average over the map coincides with the ensemble 
average (\ref{meanz}). 
\par
What we really would like to have is the correlation ergodicity
of our process. In this case, we would be able to replace the ensemble
averaging by the integration over the sphere, and to find the $K(\delta)$ 
and hence all
the $\sigma_l^2$ from a single map. Unfortunately, this is exactly what
we cannot have on a sphere. The necessary and sufficient 
condition (\ref{condergovar}) translates into the requirement
\begin{equation}   
\label{intKK}
\frac{1}{4\pi}\int _{S^2}{\rm d}\Omega |K(\rm \delta)|^2 = 0. 
\nonumber\\
\end{equation}
The left-hand-side of this equation can be calculated using Eq. (\ref{varz2})
and the orthogonality properties of the Legendre polynomials. The resulting
equation
\begin{eqnarray}
\label{sumsig}
\frac{1}{16\pi ^2}\sum _{l=1}^{\infty }(2l+1)\sigma_l^4 = 0 \nonumber 
\end{eqnarray}
can only be true if all the $\sigma_l^2$ are zero. In contrast to processes
on an infinite line or in infinite space, we do not have here an infinite 
volume factor in the denominator of the left-hand-side of Eq. (\ref{intKK})
to enable it to vanish. We are bound to do
the best what we can do with a single map - try and find out the best 
unbiased estimates for the parameters $\sigma_l^2$ and $K(\delta)$. 

\section{The best unbiased estimator for the $\sigma_l^2$}

Let us denote an estimator for the $\sigma_l^2$ by $f_l$. This is a random 
variable constructed from the original random process. One realization of
this process is the observed map. The most general 
quadratic expression for $f_l$ is given by
\begin{equation}
\label{deftildef}
f_l=\int _{S^2}\int _{S^2} {\rm d}\Omega _1 {\rm d}\Omega _2 
w_l(\vec{e}_1,\vec{e}_2)\frac{{\rm \delta }T}{T}(\vec{e}_1)
\frac{{\rm \delta }T}{T}(\vec{e}_2),
\end{equation}
where ${\rm d}\Omega=\sin \theta {\rm d}\theta {\rm d}\varphi$. 
The function $w_l(\vec{e}_1,\vec{e}_2)$ is a weight function to be 
determined from the requirements that the estimator $f_l$ is unbiased and
the minimum-variance (the best). In this formulation, the problem was 
essentially solved in Ref. \cite{Jones}. We refine and expand the 
arguments of Ref. \cite{Jones}. 
\par
An arbitrary weight function 
$w_l(\vec{e}_1,\vec{e}_2)$, being a function of two sets of angular 
coordinates, can be expanded over two sets of orthonormal spherical 
harmonics:  
\begin{eqnarray}
\label{Rd1}
w_l(\vec{e}_1,\vec{e}_2) &=
& \sum _{i=0}^{\infty }\sum _{j=0}^{\infty } d_{lij00}^{cc} 
\frac{Y_{i0}^c(\vec{e}_1)}{\sqrt{2}}\frac{Y_{j0}^c(\vec{e}_2)}{\sqrt{2}} 
\nonumber \\
&+& \sum _{i,j}\sum _{n=1}^j[d_{lijon}^{cc}
\frac{Y_{i0}^c(\vec{e}_1)}{\sqrt{2}}Y_{jn}^c(\vec{e}_2)
+d_{lijon}^{cs}\frac{Y_{i0}^c(\vec{e}_1)}{\sqrt{2}}Y_{jn}^s(\vec{e}_2)] 
\nonumber \\
&+& \sum _{i,j}\sum _{m=1}^i[d_{lijmo}^{cc}Y_{im}^c(\vec{e}_1)
\frac{Y_{j0}^c(\vec{e}_2)}{\sqrt{2}}
+d_{lijmo}^{sc}Y_{im}^s(\vec{e}_1)
\frac{Y_{j0}^c(\vec{e}_2)}{\sqrt{2}}] \nonumber \\
&+& \sum _{i,j}\sum _{m,n}[d_{lijmn}^{cc}Y_{im}^c(\vec{e}_1)Y_{jn}^c(\vec{e}_2)
+d_{lijmn}^{ss}Y_{im}^s(\vec{e}_1)Y_{jn}^s(\vec{e}_2) \nonumber \\
&+& d_{lijmn}^{cs}Y_{im}^c(\vec{e}_1)Y_{jn}^s(\vec{e}_2)
+ d_{lijmn}^{sc}Y_{im}^s(\vec{e}_1)Y_{jn}^c(\vec{e}_2)].
\end{eqnarray}
In this expression, all the coefficients $d_{lijmn}^{AB}$ ($A,B=c,s$) are real.
The weight function may be asymmetric with respect to the interchange
of $\vec{e}_1$ and $\vec{e}_2$, but the antisymmetric part of this function
will not contribute to Eq. (\ref{deftildef}) anyway. To simplify 
calculations, we
require this function to be explicitely symmetric, 
$w_l(\vec{e}_1,\vec{e}_2)=w_l(\vec{e}_2,\vec{e}_1)$. This
means that the d-coefficients must obey the relationship 
\begin{eqnarray}
\label{prop0d}
d_{lijmn}^{AB}=d_{ljinm}^{BA}. \nonumber
\end{eqnarray}
We will mostly use another representation for $w_l$ defined by:
\begin{equation}
\label{Rd2}
w_l(\vec{e}_1,\vec{e}_2)=\sum _{i=0}^{\infty }\sum _{j=0}^{\infty }
\sum _{m=-i}^i\sum _{n=-j}^j d_{lijmn}Y_{im}(\vec{e}_1)Y_{jn}^*(\vec{e}_2).
\end{equation}
In this case, the coefficients $d_{lijmn}$ are complex, and since the
function $w_l(\vec{e}_1,\vec{e}_2)$ is real, they have the property
\begin{eqnarray}
\label{prop1d}
d_{lijmn}^*=d_{lij,-m,-n}, \nonumber
\end{eqnarray}
or, if we introduce the real and imaginary parts of $d_{lijmn}$, i.e. 
$d_{lijmn}=d_{lijmn}^r+id_{lijmn}^i$,  
\begin{equation}
\label{prop2d}
d_{lijmn}^r=d_{lij,-m,-n}^r, \qquad d_{lijmn}^i=-d_{lij,-m,-n}^i.
\end{equation}
In addition, the weight function is symmetric in this representation if 
\begin{eqnarray}
\label{prop3d}
d_{lijmn}=d_{lji,-n,-m}. \nonumber
\end{eqnarray}
The link between the two representations is expressed by the equations
\begin{eqnarray}
\label{linkdd1}
d_{lij00} &=& d_{lij00}^{cc}, \nonumber \\
\label{linkdd2}
d_{lij0n} &=& \left\{ \begin{array}{ll} \frac{1}{\sqrt{2}}(d_{lij0n}^{cc}+
id_{lij0n}^{cs}) & n\ge 1 \nonumber \\
\frac{1}{\sqrt{2}}(d_{lij0,-n}^{cc}-id_{lij0,-n}^{cs}) & n\le -1, 
\end{array} \right. \nonumber \\
\label{linkdd3}
d_{lijm0} &=& \left\{ \begin{array}{ll} \frac{1}{\sqrt{2}}(d_{lijm0}^{cc}-
id_{lijm0}^{sc}) & m\ge 1 \nonumber \\
\frac{1}{\sqrt{2}}(d_{lij,-m,0}^{cc}+
id_{lij,-m,0}^{sc}) & m\le -1, \end{array} 
\right. \nonumber \\
\label{linkdd4}
d_{lijmn} &=& \left \{ \begin{array}{lll} \frac{1}{2}(d_{lijmn}^{cc}+
d_{lijmn}^{ss}-id_{lijmn}^{sc}+
id_{lijmn}^{cs}) & m \ge 1 & n\ge 1 \nonumber \\
\frac{1}{2}(d_{lij,-m,-n}^{cc}+d_{lij,-m,-n}^{ss}+id_{lij,-m,-n}^{sc}-
id_{lij,-m,-n}^{cs}) & m\le -1 & n\le -1 \nonumber \\
\frac{1}{2}(d_{lijm,-n}^{cc}-d_{lijm,-n}^{ss}-id_{lijm,-n}^{sc}-
id_{lijm,-n}^{cs}) & m\ge 1 & n\le -1 \nonumber \\
\frac{1}{2}(d_{lij,-mn}^{cc}-d_{lij,-mn}^{ss}+id_{lij,-mn}^{sc}+
id_{lij,-mn}^{cs}) & m\le -1 & n\ge 1. \end{array} \right.
\end{eqnarray}
\par
The angular integrals in Eq. (\ref{deftildef}) can be performed explicitely. 
This integration returns us from the random process $\delta T / T$ to the 
random variables - coefficients in the decomposition of 
the $\delta T / T$ - and allows us to express the estimator $f_l$ in 
terms of the general quadratic
combination of these coefficients. As a consequence, we arrive at the 
following expression for the $f_l$ in terms of the b-coefficients:
\begin{eqnarray}
\label{Esti1}
f_l &=& \sum _{i=0}^{\infty }\sum _{j=0}^{\infty }
\biggl( d_{lij00}^{cc}b_{i0}^cb_{j0}^c \nonumber \\
&+& \sum _{m=1}^i(d_{lijm0}^{cc}b_{j0}^cb_{im}^c+
d_{lijm0}^{sc}b_{j0}^cb_{im}^s)
+\sum _{n=1}^j(d_{lij0n}^{cc}b_{i0}^cb_{jn}^c+
d_{lij0n}^{cs}b_{i0}^cb_{jn}^s)
\nonumber \\
&+& \sum _{m=1}^i\sum _{n=1}^j(d_{lijmn}^{cc}b_{im}^cb_{jn}^c+
d_{lijmn}^{cs}b_{im}^cb_{jn}^s+
d_{lijmn}^{sc}b_{im}^sb_{jn}^c+d_{lijmn}^{ss}b_{im}^sb_{jn}^s)\biggr).   
\end{eqnarray}
In the representation defined by Eqs. (\ref{Rd2}), (\ref{Rc}), the last 
equation takes the form:
\begin{equation}
\label{Esti2}
f_l = \sum _{i=0}^{\infty }\sum _{j=0}^{\infty }\sum _{m=-i}^i 
\sum _{n=-j}^j d_{lijmn}c_{im}^*c_{jn}.
\end{equation}
\par
We have introduced the general expression for the estimator and can now
subject it to the desired requirements. Let us start from the mean value 
of the estimator. Using Eqs. (\ref{Esti2}) and (\ref{quadrac5}) we obtain
\begin{eqnarray}
\label{meanesti1}
\langle f_l\rangle &=& \sum _{i=0}^{\infty }\sum _{j=0}^{\infty }\sum _{m=-i}^i 
\sum _{n=-j}^j d_{lijmn}\langle c_{im}^*c_{jn}\rangle \nonumber \\
&=& \sum _{i=0}^{\infty }\sigma_i^2\sum _{m=-i}^id_{liimm}. \nonumber
\end{eqnarray}
Since the second set of equations (\ref{prop2d}) guarantees  
\begin{eqnarray}
\label{consima}
\sum _{m=-i}^id_{liimm}^i=-\sum _{m=-i}^id_{liimm}^i=0, \nonumber 
\end{eqnarray}
the mean value of $f_l$ reduces to the following manifestly real expression:
\begin{eqnarray}
\label{meanesti2}
\langle f_l\rangle=\sum _{i=0}^{\infty }\sigma_i^2\sum _{m=-i}^id_{liimm}^r. \nonumber
\end{eqnarray}
\par
We want our estimator to be unbiased, that is, we impose the
condition $\langle f_l\rangle=\sigma_l^2$. This requirement can only be achieved if 
\begin{equation}
\label{consmean}
\sum _{m=-i}^id_{liimm}^r=\delta _{li}.
\end{equation}
Equation (\ref{consmean}) form the first set of constraints on the weight
function $w_l$ and define the family of unbiased estimators.
\par 
The next step is to find, among the unbiased estimators, the one whose 
variance is minimal. The variance $\sigma_{f_l}^2$ of the random 
variable $f_l$, 
\begin{eqnarray}
\label{defvar}
\sigma _{f_l}^2 = \langle f_l^2\rangle-\langle f_l\rangle^2 , \nonumber
\end{eqnarray}
can be calculated using the definition (\ref{Esti2}) and the equations
(\ref{quarticc1}). The general expression for the variance reduces to 
the form     
\begin{equation}
\label{varesti}
\sigma _{f_l}^2 = 2\sum _{i=0}^{\infty }\sum _{j=0}^{\infty }
\sum _{m=-i}^i \sum _{n=-j}^j d_{lijmn}d_{lijmn}^*\sigma_i^2\sigma_j^2.
\end{equation}
We have to minimize this expression taking into account the constraints 
(\ref{consmean}). The expression (\ref{varesti}) is the sum of strictly
positive terms. To minimize this sum, we should set to zero as many terms
as possible. First, we need to set to zero all the d-coefficients which 
do not participate in the constraint (\ref{consmean}) and whose presence 
in the sum only increases the variance. Thus, we require to vanish all
the coefficients $d_{lijmn}^i$ and those of $d_{lijmn}^r$ which have 
indices $i\neq j$ and/or $m\neq n$. To minimize the remaining variance 
under the constraint (\ref{consmean}), we introduce the Lagrange
multipliers $\lambda _i$ and write  
\begin{eqnarray}
\label{minvar}
\delta \biggl( 2\sum _{i=0}^{\infty}\sum _{m=-i}^i(d_{liimm}^r)^2\sigma_i^4 + 
\sum _{i=0}^{\infty} \lambda _i(\sum _{m=-i}^id_{liimm}^r-
\delta _{li})\biggr)=0, \nonumber
\end{eqnarray}
Since the $d_{liimm}^r$ are treated as independent variables, the variation
of the previous expression provides us with 
\begin{equation}
\label{minre2}
4d_{liimm}^r\sigma _i^4+\lambda _i=0.
\end{equation}
The sum over $m$ of these equations together with Eq. (\ref{consmean}) 
determine the quantities $\lambda _i$:
\begin{equation}
\label{minre3}
4\sigma _i^4\delta _{li}+(2i+1)\lambda _i=0.
\end{equation}
Using the expression (\ref{minre3}) for the Lagrange multipliers $\lambda _i$ 
in Eq. (\ref{minre2}) allows us to write 
\begin{eqnarray}
\label{bestd^r}
d_{liimm}^r=\frac{1}{2i+1}\delta _{li}.  \nonumber
\end{eqnarray}
Thus, taking into account all relationships, we obtain the complete set of
constraints on the weight function (\ref{Rd2}) which make the estimator 
$f_l$ unbiased and best: 
\begin{equation}
\label{bestesti}
d_{lijmn}^r=\frac{1}{2i+1}\delta _{li}\delta _{ij}\delta _{mn}, 
\qquad d_{lijmn}^i=0.
\end{equation}
In the representation (\ref{Rd1}), this amounts to 
\begin{eqnarray}
\label{bestcan1}
d_{lij00}^{cc} &=& \frac{1}{2i+1}\delta _{li}\delta _{ij}, \\
\label{bestcan2}
d_{lijmn}^{cc} &=& d_{lijmn}^{ss}=
\frac{1}{2i+1}\delta _{li}\delta _{ij}\delta _{mn},
\end{eqnarray}
with all other coefficients being zero. 
\par
Having found all the $d$-coefficients, we can write the weight function
$w_l(\vec{e}_1,\vec{e}_2)$ explicitely. Using the definition (\ref{Rd2}) and
the found expressions (\ref{bestesti}) we obtain
\begin{eqnarray}
\label{bestwl1}
w_l(\vec{e}_1,\vec{e}_2) &=& \sum _{i=0}^{\infty }\sum _{j=0}^{\infty }
\sum _{m=-i}^i\sum _{n=-j}^j \frac{1}{2i+1}\delta _{li}\delta _{ij}\delta _{mn}
Y_{im}(\vec{e}_1)Y_{jn}^*(\vec{e}_2) \nonumber \\
\label{bestwl2}
&=& \frac{1}{2l+1}\sum _{m=-l}^l Y_{lm}(\vec{e}_1)Y_{lm}^*(\vec{e}_2) \\
\label{bestwl3}
&=& \frac{1}{4\pi }P_l(\cos \delta _{12}),
\end{eqnarray}
where, in the last step, the summation theorem for spherical harmonics 
has been used. Therefore, the best unbiased estimator for $\sigma_l^2$ 
can be written as:
\begin{equation}
\label{bestftilde}
f_l=\frac{1}{4\pi }\int _{S^2}\int _{S^2} {\rm d}\Omega _1{\rm d}\Omega _2
P_l(\cos \delta _{12}) \frac{{\rm \delta }T}{T}(\vec{e}_1)
\frac{{\rm \delta }T}{T}(\vec{e}_2).
\end{equation}
This formula answers the question what to do with a given map in order to get
a concrete number - the best substitute for the true parameter $\sigma_l^2$. 
The answer is to perform with the map the integrations prescribed by this
formula. In fact, the integrations can be further simplified.    
\par
The estimator (\ref{bestftilde}) contains a double integral of the product
of two functions $\delta T / T$ and therefore can be called a quadratic
estimator. However, this estimator can be presented as a product of two
linear estimators, i.e. a product of two appropriate single integrals of 
the function $\delta T / T$. Indeed, using the summation 
theorem, Eqs. (\ref{bestwl2}), (\ref{bestwl3}), formula (\ref{bestftilde})  
can be written as
\begin{eqnarray}
\label{estisquare}
f_l=\frac{1}{2l+1}\sum _{m=-l}^l\int_{S^2} {\rm d}\Omega _1 
Y_{lm}(\vec{e}_1)\frac{{\rm \delta}T}{T}(\vec{e}_1)
\int_{S^2} {\rm d}\Omega _2 
Y_{lm}^*(\vec{e}_2)\frac{{\rm \delta }T}{T}(\vec{e}_2). \nonumber  
\end{eqnarray}
This formula shows that it is sufficient to perform one appropriately
weighted integration over the sphere with further multiplications and
summations. Moreover, the remaining integrals define the $c_{lm}$ 
coefficients. So, we obtain the following expression for the best 
unbiased estimator in terms of the original random coefficients:  
\begin{equation}
\label{esticl2}
f_l=\frac{1}{2l+1}\sum _{m=-l}^lc_{lm}^*c_{lm}=\frac{c_l^2}{2l+1}.
\end{equation} 
Of course, this is the same expression which could be obtained by
inserting Eq. (\ref{bestesti}) into Eq. (\ref{Esti2}) or by inserting 
Eqs. (\ref{bestcan1}), (\ref{bestcan2}) into Eq. (\ref{Esti1}).
\par
We have found the estimator with the smallest possible variance among
all unbiased estimators. It is useful to write this minimal variance 
explicitely. This can be found either from Eqs. (\ref{varesti}) 
and (\ref{bestesti}) or from Eqs. (\ref{esticl2}) and 
(\ref{varbc^2}). The result is       
\begin{eqnarray}
\label{varbestesti}
\sigma _{f_l}^2=\frac{2}{2l+1}\sigma_l^4. \nonumber
\end{eqnarray}

\section{The best unbiased estimator for the $K(\delta)$}

The best unbiased estimator for $\sigma_l^2$ is also the best
unbiased estimator for the multipole moments $C_l$ of the correlation 
function $K(\delta)$, see Eqs. (\ref{varz2}), (\ref{corresCsig}). Since the
parameter $K(\delta)$ is a combination of the parameters $\sigma_l^2$, it is
not surprising that the best unbiased estimator for $K(\delta)$ turns out
to be the same combination of the best unbiased 
estimators for $\sigma_l^2$. It is interesting and instructive 
to follow this relationship in detail.  
\par
Let us denote an estimator of the $K(\delta)$ by $f(\delta)$. This is a random
variable constructed from the random process $\delta T / T$. The most 
general (quadratic) expression for the $f(\delta)$ can be written as    
\begin{equation}
\label{defvtilde1}
f(\delta ) = \int _{S^2}\int _{S^2} {\rm d}\Omega _1 {\rm d}\Omega _2 
w(\vec{e}_1, \vec{e}_2, \delta) 
\frac{{\rm \delta }T}{T}(\vec{e}_1)\frac{{\rm \delta }T}{T}(\vec{e}_2), 
\end{equation}
where the $\delta$ is a fixed angle, whereas the angle between variable
directions $\vec{e}_1$ and $\vec{e}_2$ will be denoted $\delta _{12}$. 
The arbitrary weight function $w(\vec{e}_1, \vec{e}_2, \delta)$ can be
expanded, without loss of generality, over the Legendre polynomials 
$P_l(\cos \delta)$:
\begin{eqnarray}
\label{ww_l}
w(\vec{e}_1, \vec{e}_2, \delta)=\frac{1}{4\pi }\sum _{l=0}^{\infty }
(2l+1)P_l(\cos \delta )w_l(\vec{e}_1, \vec{e}_2). \nonumber
\end{eqnarray}
The estimator $f(\delta)$ takes the form 
\begin{eqnarray}  
\label{defvtilde2}
f(\delta) &=& \frac{1}{4\pi }\sum _{l=0}^{\infty}(2l+1)P_l(\cos \delta )
\int _{S^2}\int _{S^2} {\rm d}\Omega _1 {\rm d}\Omega _2 
w_l(\vec{e}_1, \vec{e}_2) \frac{{\rm \delta }T}{T}(\vec{e}_1)
\frac{{\rm \delta }T}{T}(\vec{e}_2) \nonumber \\
\label{defvtilde3}
&=& \frac{1}{4\pi }\sum _{l=0}^{\infty }(2l+1)f_lP_l(\cos \delta ).
\nonumber
\end{eqnarray}
\par
Now, we want the $f(\delta)$ to be unbiased and best estimator of $K(\delta)$.
The estimator $f(\delta)$ is unbiased if $\langle f_l\rangle = \sigma_l^2$, and it is
the best if the variance of the $f_l$ is the smallest one. In other words,
we return to the solved problem (Section 5) for the estimator $f_l$.
Using the results of the previous Section, we can write for the best
unbiased estimator of the $K(\delta)$:
\begin{equation}
\label{bestvtilde}
f(\delta )=\frac{1}{4\pi }\sum _{l=0}^{\infty }P_l(\cos \delta )c_l^2.
\end{equation}
The weight function $w(\vec{e}_1, \vec{e}_2, \delta)$ is given, taking 
into account Eq. (\ref{bestwl3}), by
\begin{equation}
\label{bestw}
w(\vec{e}_1, \vec{e}_2, \delta)=
\frac{1}{(4\pi )^2}\sum _{l=0}^{\infty } (2l+1)P_l(\cos \delta )
P_l(\cos \delta _{12}).
\end{equation}
This last equation can also be written as
\begin{eqnarray}
\label{coefw2}
w(\vec{e}_1, \vec{e}_2, \delta)=
\frac{1}{4\pi }\sum _{l=0}^{\infty }\sum _{m=-l}^lP_l(\cos \delta )
Y_{lm}(\vec{e}_1)Y_{lm}^*(\vec{e}_2), \nonumber
\end{eqnarray}
showing that the double integral in Eq. (\ref{defvtilde1}) decays into
the products and summations of the appropriately weighted single integrals.
This form of the weight function permits an immediate recovery of the 
already known result (\ref{bestvtilde}). The variance of the best 
estimator (\ref{bestvtilde}) is:
\begin{eqnarray}
\label{varv}
\sigma _{f(\delta)}^2=\frac{1}{8\pi ^2}\sum _{l=0}^{\infty }
(2l+1) \sigma _l^4 P_l^2(\cos \delta ). \nonumber
\end{eqnarray}
\par
The derived formulas answer the question what to do with a given map
in order to derive the best unbiased estimate for the correlation
function $K(\delta)$. The outlined prescription essentially goes through
the derivation of the best unbiased estimate for $\sigma_l^2$. However,
the weight function (\ref{bestw}) allows also a different procedure
for the derivation of the estimator and the estimate: the 
direct integration of the map, but with the help of the $\delta$-function. 
\par
Let us denote $\cos \delta = x$ and $\cos \delta_{12} = x_0$. Let us 
present (define) the $\delta$-function $\delta(x-x_{0})$ as an
expansion over the Legendre polynomials:
\begin{equation}
\label{deltaf}
\delta(x-x_{0}) = \sum_{l=0}^{\infty} a_l P_l(x).
\end{equation}
To find the coefficients $a_l$, multiply both sides of Eq. (\ref{deltaf})  
by $P_{m}(x)$ and integrate by $x$ from $-1$ to $1$. The result is
\begin{eqnarray}
\label{a_l}
a_l = \frac{2l+1}{2}P_{l}(x_0) \nonumber
\end{eqnarray}  
and
\begin{eqnarray}
\label{taylordelta}
\delta(x-x_0) = \sum_{l=0}^{\infty} \frac{2l+1}{2}P_{l}(x)P_{l}(x_0).
\nonumber
\end{eqnarray}
Thus, the weight function (\ref{bestw}) can be written as
\begin{equation}
\label{wdelta}
w (\vec{e}_1,\vec{e}_2,\delta)= 
\frac {1}{8 \pi^2} \delta (\cos \delta - \cos \delta_{12}).
\end{equation}   
\par
Let us show that the integration in Eq. (\ref{defvtilde1}) with the
weight function (\ref{wdelta}) does indeed provide us with the same
result (\ref{bestvtilde}). The product of two $\delta T/ T$ is the
random process   
$v\equiv \frac{{\rm \delta }T}{T}(\vec{e}_1)
\frac{{\rm \delta }T}{T}(\vec{e}_2)$. We have access to one realization 
of this process. To integrate the $v$ over all directions 
$\vec{e}_1, \vec{e}_2$ on the sky separated by a fixed angle $\delta$ 
one can 
proceed as follows. At the first step, rotate the vector $\vec{e}_2$ around 
the fixed direction defined by $\vec{e}_1$ and integrate the $v$ over the 
circle traced by the vector $\vec{e}_2$ on the sphere. The result 
will depend only on $(\theta _1,\varphi _1)$ - the coordinates of the 
vector $\vec{e}_1$. At the second step, integrate the result over 
all $\theta _1$ and $\varphi _1$, letting the vector $\vec{e}_1$ to run over 
the whole sphere. The final result, taking into account also
the factor $1/8 \pi^2$ in (\ref{wdelta}), should be the random variable we 
are interested in. Let us do this computation in practice.
\par
Every function on a sphere can be expanded in the basis of spherical harmonics.
In particular, the function $Y_{lm}(\theta ,\varphi )$ can be expanded in 
the basis $\{Y_{lr}(\theta ',\varphi '); l\ge 0, -l\le r\le l\}$ according 
to the formula:
\begin{eqnarray}
\label{expanY}
Y_{lm}(\theta ',\varphi ')=
\sum _{r=-l}^lY_{lr}(\theta ,\varphi )D_{rm}^l(\alpha ,\beta ,\gamma ).
\nonumber
\end{eqnarray}
The coefficients of this expansion are called the Wigner D-functions 
\cite{VMK}.
They depend on the Euler angles $\alpha $, $\beta $, $\gamma $ describing 
the rotation which transforms the direction $(\theta ,\varphi )$ into the 
direction $(\theta ',\varphi ')$. Since the rotation specified by the 
angles $\alpha $, $\beta =-\theta _1$, $\gamma =-\varphi _1$ brings the 
vector pointing out to the north pole to the direction defined by 
$(\theta _1,\varphi _1)$, we can write:
\begin{equation}
\label{defY1}
Y_{lm}(\vec{e}_1)=\sum _{r=-l}^lY_{lr}(0,-)D_{rm}^l(\alpha, -\theta _1, 
-\varphi _1).
\end{equation}
In the same manner, the $Y_{pq}(\vec{e}_2)$ can be expressed as
\begin{equation}
\label{defY2}
Y_{pq}(\vec{e}_2)=\sum _{s=-p}^pY_{ps}(\delta ,\chi )
D_{sq}^p(\alpha ,-\theta _1, -\varphi _1).
\end{equation}
Indeed, when the vector $\vec{e}_1$ points out to the north pole, 
the $\theta $-coordinate of $\vec{e}_2$ is simply $\delta $. 
\par
Using the definition (\ref{Rc}) and the formulae (\ref{defY1}),
(\ref{defY2}), we can present the random process $v$ as
\begin{eqnarray}
\label{expanv}
v=\sum _{lm}\sum _{pq}\sum _{rs}c_{lm}c_{pq}Y_{lr}(0,-)Y_{ps}(\delta ,\chi )
D_{rm}^l(\alpha ,-\theta _1,-\varphi _1)
D_{sq}^p(\alpha ,-\theta _1,-\varphi _1). \nonumber
\end{eqnarray}
Let us now perform the two step integration procedure described above. 
The first step amounts to the integration of $v$ over the angle $\chi $ 
from $0$ to $2\pi $. Using the explicit form of the spherical harmonics 
given in the Appendix, we get:
\begin{eqnarray}
\label{intchi}
& & \int _0^{2\pi } v {\rm d}\chi = \sum _{lm}\sum _{pq}\sum _{rs} 
c_{lm}c_{pq}(\frac{2l+1}{4\pi })^{1/2}\delta _{r0}
(\frac{2p+1}{4\pi })^{1/2}\biggl[\frac{(p-s)!}{(p+s)!}\biggr] \times 
\nonumber \\
& & P_{ps}(\cos \delta )D_{rm}^l(\alpha ,-\theta _1,-\varphi _1)
D_{sq}^p(\alpha ,-\theta _1,-\varphi _1)
\int _0^{2\pi } e^{is\chi} {\rm d}\chi. \nonumber
\end{eqnarray}
In the last expression, the integral 
$\int _0^{2\pi } e^{is\chi}{\rm d}\chi$ is simply $2\pi \delta _{s0}$. 
Using the relationship 
\begin{eqnarray} 
\label{linkDY}
\sqrt {\frac{2l+1}{4\pi}}D_{om}^l(\alpha ,\beta ,\gamma )=
Y_{lm}(\beta ,\gamma ), \nonumber
\end{eqnarray}
we obtain 
\begin{eqnarray}
\label{intchi2}
\int _0^{2\pi } v {\rm d}\chi = 2 \pi \sum _{lm}\sum _{pq}c_{lm}c_{pq}
P_p(\cos \delta )Y_{lm}(\theta _1,\varphi _1)Y_{pq}(\theta _1,\varphi _1).
\nonumber
\end{eqnarray}
As expected, the result depends only on $(\theta _1, \varphi _1)$. 
In order to complete the procedure, we have to integrate this result 
over $\Omega _1$. We find
\begin{eqnarray}
\label{intomega}
\int {\rm d}\Omega _1\int _0^{2\pi } v {\rm d}\chi  
&=& 2\pi \sum _{lm}\sum _{pq}c_{lm}c_{pq}P_p(\cos \delta )\int {\rm d}\Omega _1
Y_{lm}(\theta _1,\varphi _1)Y_{pq}(\theta _1,\varphi _1) \nonumber\\
\label{intomega2}
&=& 2\pi \sum _{l=0}^{\infty }P_l(\cos \delta )c_l^2. \nonumber
\end{eqnarray}
Restoring the factor $1/8 \pi^2$ from (\ref{wdelta}), we arrive at   
the best unbiased estimator (\ref{bestvtilde}). The same procedure 
performed over a given map provides us with the best unbiased estimate
of the correlation function $K(\delta)$.  

\section{Best estimates and the error bars}

We will denote the best unbiased estimate for the parameter $\sigma_l^2$ 
by $(\sigma_l^2)_{BU}$. Along with this one, there
exist other estimates of the same parameter, for example, the maximum-
likelihood estimate $(\sigma_l^2)_{ML}$. Not surprisingly, for the 
postulated distributions (\ref{pdfb1}), (\ref{pdfb2}), (\ref{pdfb3}) 
these estimates coincide.
\par
Apparently the most "naive" evaluation of the $\sigma_l^2$, giving
nevertheless the correct result, would be the following one. From a given
map one derives the set of the observed coefficients $b_{lm}^{(map)}$ 
(that is to say, the set $\{b_{l0}^c,b_{lm}^c,b_{lm}^s\}$, $m \ge 1$). We know
(postulate) that each of them is drawn from the normal zero-mean distribution
\begin{equation}
\label{pdfblm}
f(b_{lm})=\frac{1}{\sqrt{2\pi }\sigma _l}e^{-\frac{b_{lm}^2}{2\sigma _l^2}}. 
\end{equation}
Each of the observed $b_{lm}$ coefficients can be used for 
the maximum-likelihood evaluation of the corresponding $\sigma_l^2$. 
[We omit the label (map) when it is clear that we deal with the observed
quantities.] One finds this estimate, denoted $\sigma_{l(m)}^2$, by 
assuming that 
the p.d.f. (\ref{pdfblm}) reaches its maximum at the observed $b_{lm}$.
The result is known: $\sigma_{l(m)}^2 = b_{lm}^2$. Indeed,
\begin{eqnarray}  
\label{lnf}
\ln f = -{\frac {1}{2}} \ln (\sigma_l^2) - \frac{b_{lm}^2}{2\sigma_l^2} + C,
\nonumber
\end{eqnarray}
where $C$ is a constant. Then, one gets 
\begin{eqnarray}
\label{dlnf}
\frac{{\rm d} \ln f}{{\rm d} \sigma_l^2} = -{\frac {1}{2 \sigma_l^2}}+
\frac{b_{lm}^2}{2\sigma_l^4}, \nonumber
\end{eqnarray}
and the condition
\begin{equation}
\label{dlnfz}
\frac{{\rm d} \ln f}{{\rm d} \sigma_l^2} = 0 
\end{equation}
leads to the stated result. Since for every $l$ we have $2l + 1$ independent
$b_{lm}$ coefficients and, hence, $2l + 1$ independent evaluations, 
the estimate of the true $\sigma_l^2$ is given by  
\begin{eqnarray}  
\label{MLesti}
\frac{1}{2l+1} \sum _{m} \sigma_{l(m)}^2=
\frac{1}{2l+1}\sum _{m}b_{lm}^2=\frac{b_l^2}{2l+1}. \nonumber
\end{eqnarray} 
This number coincides with the $(\sigma_l^2)_{BU}$ determined 
by Eq. (\ref{esticl2}).   
\par
A similar maximum-likelihood evaluation of $\sigma_l^2$ is based on the
joint p.d.f. for all coefficients $b_{lm}$ (with the same index $l$) which
is simply the product of the individual p.d.f.'s (\ref{pdfblm}):       
\begin{eqnarray}
\label{pdfblmj}
f(b_{l0}^c,b_{lm}^c,b_{lm}^s)=\frac{1}{(2\pi\sigma _l^2)^{(2l+1)/2}}
e^{-\frac{b_l^2}{2\sigma _l^2}}. \nonumber
\end{eqnarray}
Imposing the condition (\ref{dlnfz}), we arrive at the same result
\begin{eqnarray}
\label{sigml}
(\sigma_l^2)_{ML}=\frac{b_l^2}{2l+1}=(\sigma_l^2)_{BU}. \nonumber 
\end{eqnarray}
\par
Finally, we can give the maximum-likelihood estimation based on the
p.d.f. (\ref{pdfbl2}) for the quantities $b_l^2$. Repeating the same steps, 
we come again to the result  
\begin{equation}
\label{sigmln}
(\sigma_l^2)_{ML}=\frac{b_l^2}{n}.
\end{equation}
\par
Thus, in the first approximation, we can write for the true value of
$\sigma_l^2$: $\sigma_l^2 = (\sigma_l^2)_{BU}$. In the next approximation,
we want to make this statement more accurate by assigning the error bars.
This is the matter of definitions, and there are many ways of doing this. We
will use the distribution function (\ref{pdfbl2}). This function
attains its maximum at the measured $b_l^2$ and the 
estimated $(\sigma_l^2)_{ML}$, Eq. (\ref{sigmln}). The value of the $f$ at the
maximum is   
\begin{eqnarray}
\label{fmax}
f_{max} = \frac{(b_l^2)^{-1} e^{-{n/2}}}{(n/2 - 1)! (2/n)^{n/2}}. 
\nonumber
\end{eqnarray}
When the $\sigma_l^2$, treated as a variable parameter, deviates from
$(\sigma_l^2)_{ML}$, the value of the $f$ decreases as compared with the
$f_{max}$. We establish the error bars for $\sigma_l^2$ by requiring that
the value of $f$ does not drop below some confidence level
\begin{equation}
\label{feq}
f = k f_{max},
\end{equation}
where $k$, $k < 1$, is a fixed number. Within the error box are included
all $\sigma_l^2$ surrounding $(\sigma_l^2)_{ML}$ and up to the 
boundaries $(\sigma_l^2)_k$ determined by two solutions to the
equation (\ref{feq}).      
\par
Let us denote $x \equiv (\sigma_l^2)_{ML}/(\sigma_l^2)_k$. Equation 
(\ref{feq}) takes the form
\begin{equation}  
\label{eqx}
\ln x - x = \frac{2}{n} \ln k - 1.
\end{equation} 
Obviously, $x = 1$ if $k = 1$. Let us now consider small deviations from
this solution for $k < 1$. We write $x = 1 - y$, where $|y| << 1$. By
expanding the $\ln x$ in terms of $y$ and considering the first nonvanishing 
approximation to the equation (\ref{eqx}), we find $y^2 = - (4/n) \ln k$. 
That is, the two wanted solutions are    
\begin{eqnarray}
\label{y}
y = \pm \frac{2\sqrt{-\ln k}}{\sqrt{n}}. \nonumber
\end{eqnarray}
The condition of their applicability is $-(4/n)\ln k << 1$. Thus, 
in this approximation, we can write the true $\sigma_l^2$ as 
\begin{equation}
\label{siger}
\sigma_l^2 = (\sigma_l^2)_{BU}\bigl(1 \pm \frac{2\sqrt{-\ln k}}
{\sqrt {2l + 1}}\bigr). 
\end{equation}
\par
The choice of $k$ is in our hands. If the distribution $f(z)$ were a normal
zero-mean distribution, then a reasonable choice of $k$ would be 
$k = e^{-1/2}$, because $f(z=\sigma) = e^{-1/2} f_{max}$. The
$\chi^2$ distribution (\ref{pdfbl2}) is not a normal one, but approaches a 
normal distribution for large values of $l$. As a guidance, we will use    
$k = e^{-1/2}$ in Eq. (\ref{siger}). Then we get     
\begin{eqnarray}
\label{sigern}
\sigma_l^2 = (\sigma_l^2)_{BU}\bigl(1 \pm \sqrt{ \frac{2}{2l + 1}}\bigr). 
\nonumber
\end{eqnarray}
\par
This formula becomes progressively inaccurate for small $l$. Specifically
for $l = 2$ this formula would imply the error at the level $\pm 0.6$. 
However, a direct derivation of the error bars from equation (\ref{eqx}) (and 
assuming $k = e^{-1/2}$ ) gives     
\begin{eqnarray}
\label{sigl2}
\sigma_2^2 = (\sigma_2^2)_{BU}(1 + \sigma),  \nonumber
\end{eqnarray}
where $\sigma$ lies between $+ 1.0$ and $-0.4$.
Note the assymetry of the error interval: the larger than 
$(\sigma_2^2)_{BU}$ values are more tolerable than the smaller ones.  

\section{Acknowledgments}

L. G. appreciates very useful conversations with Serge Reynaud,
Don Page, and Matt Visser. J. M. would like to thank Richard Kerner 
for constant encouragements and Nathalie Deruelle for having raised 
his interest in the problem of ergodicity during conversations at 
DAMTP (Cambridge).

\section{Appendix}

The complex spherical harmonics $Y_{lm}(\theta ,\varphi )$ are defined 
by the expression:
\begin{eqnarray}
\label{defY}
Y_{lm}(\theta ,\varphi )=\frac{1}{\sqrt{2}}\biggl(\frac{2l+1}{2\pi }
\frac{(l-|m|)!}{(l+|m|)!}\biggr)^{1/2}P_{l|m|}(\cos \theta )e^{im\varphi }.
\nonumber
\end{eqnarray}
In this expression, $l\ge 0$ and $-l\le m\le l$. The functions 
$Y_{lm}(\theta ,\varphi)$ satisfy the relationship $Y_{lm}^*=Y_{l,-m}$. 
On the other hand, the real spherical harmonics are defined by the 
equations:
\begin{eqnarray}
\label{defYc}
Y_{lm}^c(\theta ,\varphi ) &=& \biggl(\frac{2l+1}{2\pi }
\frac{(l-m)!}{(l+m)!}\biggr)^{1/2}P_{lm}(\cos \theta )\cos m\varphi ,
\nonumber \\
\label{defYs}
Y_{lm}^s(\theta ,\varphi ) &=& \biggl(\frac{2l+1}{2\pi }
\frac{(l-m)!}{(l+m)!}\biggr)^{1/2}P_{lm}(\cos \theta )\sin m\varphi ,
\nonumber
\end{eqnarray}
where $l\ge 0$ but $0\le m\le l$. The indices $c$, $s$ indicate the 
presence of $\cos m\varphi$ or $\sin m\varphi$, respectively. The 
link between the complex and real spherical harmonics is given by
\begin{eqnarray}
\label{linkYYc}
Y_{lm}^c &=& \frac{1}{\sqrt{2}}(Y_{lm}+Y_{lm}^*) \quad m\ge 0, \nonumber \\
\label{linkYYs} 
Y_{lm}^s &=& \frac{1}{i\sqrt{2}}(Y_{lm}-Y_{lm}^*) \quad m\ge 0 . \nonumber
\end{eqnarray}
The scalar product of two functions $f(\theta ,\varphi )$ and 
$g(\theta ,\varphi )$ on the sphere is defined by
\begin{eqnarray}
\label{defsp}
(f,g)=\int _0^{2\pi }{\rm d}\varphi \int _0^{\pi }{\rm d}\theta \sin \theta
f^*(\theta ,\varphi )g(\theta ,\varphi ). \nonumber
\end{eqnarray}
Then, we have the following properties:
\begin{eqnarray}
\label{spY}
(Y_{lm},Y_{l'm'}) &=& \delta _{ll'}\delta _{mm'} \quad -l\le m\le l, 
\nonumber \\
\label{spYc0}
(\frac{Y_{l0}^c}{\sqrt{2}}, \frac{Y_{l'0}^c}{\sqrt{2}}) &=& \delta _{ll'}, 
\nonumber \\
\label{spYcs}
(Y_{lm}^A,Y_{l'm'}^B) &=& \delta _{ll'}\delta _{mm'}\delta ^{AB} 
\quad m\ge 1 \quad A,B=c,s. \nonumber
\end{eqnarray}
The functions $\frac{Y_{l0}^c}{\sqrt{2}}$ $(l \ge 0)$, $Y_{lm}^c$ $(l \ge 1,
l \ge m \ge 1)$, $Y_{lm}^s$ $(l \ge 1, l \ge m \ge 1)$ form a complete
orthonormal basis.

\end{document}